\documentclass[pra,twocolumn,amsmath,amssymb,superscriptaddress]{revtex4-2}
\usepackage{graphicx,amsmath,relsize,epstopdf,color,mathtools,bm,newtxtext,newtxmath,braket,rotating}
\usepackage[hyphenbreaks]{breakurl}
\usepackage[colorlinks=true,linkcolor=blue,citecolor=blue,urlcolor =blue]{hyperref}
\usepackage[normalem]{ulem}
\usepackage[table,xcdraw]{xcolor}
\usepackage{braket}

\usepackage{soul,xcolor}
\begin{document}
\title{Protecting Quantum Modes in Optical Fibers}

\author{M. A. T. Butt}
\affiliation{Max-Planck-Institut f\"{u}r die Physik des Lichts, 91058 Erlangen, Germany}
\affiliation{Institut f\"ur Optik, Information und Photonik,
University Erlangen-Nuremberg, 91058 Erlangen, Germany}
\affiliation{School of Advanced Optical Technologies,
University Erlangen-Nuremberg, 91052 Erlangen, Germany}
\affiliation{College of Aeronautical Engineering (CAE),  National University of Sciences and Technology, 24090 Risalpur, Pakistan}

\author{P. Roth}
\affiliation{Max-Planck-Institut f\"{u}r die Physik des Lichts, 91058 Erlangen, Germany}
\affiliation{Institut f\"ur Optik, Information und Photonik,
University Erlangen-Nuremberg, 91058 Erlangen, Germany}

\author{G. K. L. Wong}
\affiliation{Max-Planck-Institut f\"{u}r die Physik des Lichts, 91058 Erlangen, Germany}
\affiliation{Institut f\"ur Optik, Information und Photonik,
University Erlangen-Nuremberg, 91058 Erlangen, Germany}

\author{M. H. Frosz}
\affiliation{Max-Planck-Institut f\"{u}r die Physik des Lichts, 91058 Erlangen, Germany}

\author{L.~L. S\'{a}nchez-Soto}
\affiliation{Max-Planck-Institut f\"{u}r die Physik des Lichts, 91058 Erlangen, Germany}
\affiliation{Institut f\"ur Optik, Information und Photonik,
University Erlangen-Nuremberg, 91058 Erlangen, Germany}
\affiliation{Departamento de \'Optica, Facultad de F\'{\i}sica,
Universidad Complutense, 28040 Madrid, Spain}

\author{E.~A.~Anashkina}
\affiliation{A.V. Gaponov-Grekhov Institute of Applied Physics, Russian Academy of Sciences, 603950 Nizhny Novgorod, Russia}
\affiliation{Lobachevsky State University of Nizhny Novgorod, 603022, Nizhny Novgorod, Russia}

\author{A. V. Andrianov}
\affiliation{A.V. Gaponov-Grekhov Institute of Applied Physics, Russian Academy of Sciences, 603950 Nizhny Novgorod, Russia}

\author{P. Banzer}
\affiliation{Max-Planck-Institut f\"{u}r die Physik des Lichts, 91058 Erlangen, Germany}
\affiliation{Institut f\"ur Optik, Information und Photonik, University Erlangen-Nuremberg, 91058 Erlangen, Germany}
\affiliation{School of Advanced Optical Technologies, University Erlangen-Nuremberg, 91052 Erlangen, Germany}
\affiliation{Institute of Physics, University of Graz, 8010 Graz, Austria}

\author{P. St.J. Russell}
\affiliation{Max-Planck-Institut f\"{u}r die Physik des Lichts, 91058 Erlangen, Germany}

\author{G.~Leuchs}
\affiliation{Max-Planck-Institut f\"{u}r die Physik des Lichts, 91058 Erlangen, Germany}
\affiliation{Institut f\"ur Optik, Information und Photonik,
University Erlangen-Nuremberg, 91058 Erlangen, Germany}
\affiliation{Institute of Applied Physics, Russian Academy of Sciences, 603950 Nizhny Novgorod, Russia}

\begin{abstract}
Polarization-preserving fibers maintain the two polarization states of an orthogonal basis. Quantum communication, however, requires sending at least two nonorthogonal states and these cannot both be preserved.  We present a new scheme that allows for using polarization encoding in a fiber not only in the discrete, but also in the continuous-variable regime. For the example of a helically twisted photonic-crystal fibre, we experimentally demonstrate that using appropriate nonorthogonal modes, the polarization-preserving fiber does not fully scramble these modes over the full Poincar\'e sphere, but that the output polarization will stay on a great circle; that is, within a one-dimensional protected subspace, which can be parametrized by a single variable. This will allow for more efficient measurements of quantum excitations in nonorthogonal modes.
\end{abstract}

\maketitle

\section{Introduction}

Modern communication technology based on light as the information carrier has proven to be an invaluable tool for transmitting large amounts of data at high speeds~\cite{Ekert:1991wo,Gisin:2007vl,Kimble:2008uv}. The effective transfer of this information is achieved by modulating some of the parameters of the radiation, most often amplitude or phase. But the unavoidable vector nature of light cannot be neglected and, in practice, can also affect the performance. For example, in fiber-based systems~\cite{Arumugam:2001tx}, small stress-induced  birefringence varying along the fiber leads to polarization mode dispersion (PMD)~\cite{Gordon:2000uf,Galtarossa:2006th,Gordon:2000uf}, which ultimately limits the data rate~\cite{Essiambre:2010wd}.  

\begin{figure*}[t]
  \centerline{\includegraphics[width=1.53\columnwidth]{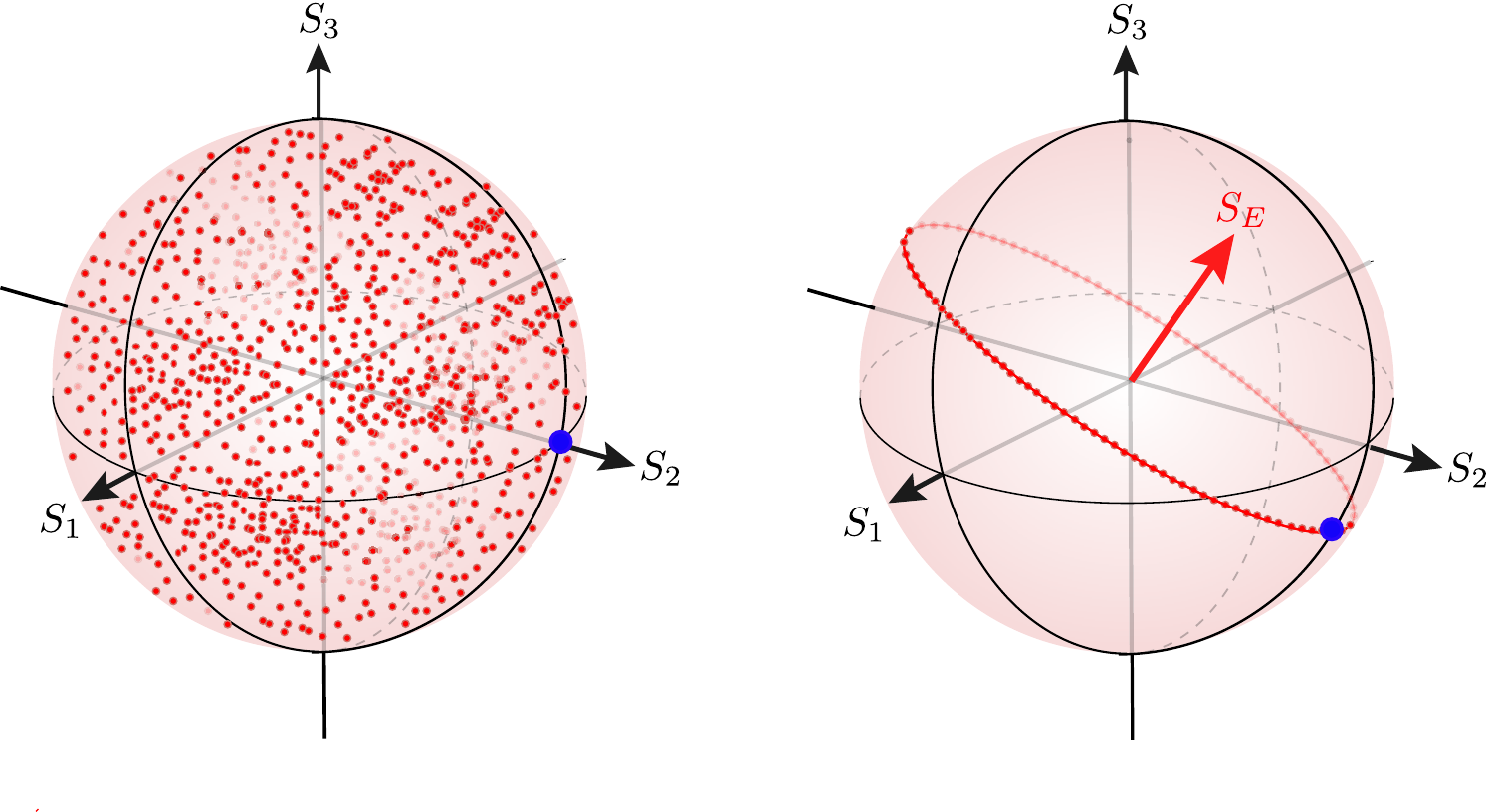}}
  \caption{Polarization evolution in an optical fiber plotted on the Poincar\'e sphere. Left panel: standard fiber, where PMD leads to wandering of the state of light over the whole sphere. Right panel: PMF with axis $S_E$ is defined by the eigenmodes of the  fiber. The evolution occurs along the circle perpendicular to $S_E$ containing the initial state.}
  \label{fig1}
\end{figure*}

For a standard fiber, PMD leads to a wandering of the excited mode of light over the whole Poincar\'e sphere~\cite{Brosseau:1998aa,Goldstein:2011aa,Gil:2016aa,Goldberg:2021tx}, so that after a sufficiently long propagation the probability of finding the mode  at a particular point is equally distributed over the whole sphere, as sketched in Fig.~\ref{fig1}. This is usually called polarization scrambling. To overcome this problem, polarizing-maintaining fibers (PMFs) were developed~\cite{Noda:1986wv}. They preserve two polarization eigenmodes, which defines the axis $S_E$ of the Poincar\'e sphere. In this way, the polarization mode is scrambled only along a circle perpendicular to $S_E$. 

Most commonly, PMFs preserve linear polarization by having largely differing propagation speeds for two orthogonal polarization modes, so that there is no cross-coupling. In this way, one gets two different information channels for classical communication without limitation by PMD. PMFs that preserve circular polarization work similarly, although these fibers are considerably more difficult to obtain. 

In quantum communication, the situation is different. An essential ingredient is the possibility to discriminate, in the simplest case, two nonorthogonal states~\cite{Chuang:2000fk}. These could be nonorthogonal modes in amplitude, phase or polarization. The corresponding type of measurement is relevant for important tasks in, e.g., quantum key distribution (QKD)~\cite{Xu:2020tu,Pirandola:2020vs} and in entanglement-swapping protocols~\cite{Gisin:2002wz}. Since amplitude modulation is generally less reliable and phase modulation requires a very stable local oscillator to recover the signal,  polarization modulation is often the preferred choice~\cite{Heim:2014wn}. Although a variety of techniques are at hand, one has to demodulate the signal, which requires measuring all three Stokes parameters. Determining the three Stokes parameters introduces two units of quantum noise  due to the attempt to measure noncommuting variables simultaneously~\cite{Arthurs:1965aa,Stenholm:1992aa}.  Thus, searching the whole sphere complicates the detection by introducing an unavoidable quantum penalty~\cite{Goldberg:2022aa,Feng:2004aa}.

One may think of resorting to a PMF to avoid this extra noise. If one uses a linear PMF (with axis $S_E=S_1$), and if one chooses the initial state on the great circle defined by $S_E$, then estimating the signal requires only measuring in a certain linear combination of $S_2$ and $S_3$, that is elliptical polarization in general. If one had, however, a fiber preserving circular polarization (with axis $S_E=S_3$), then measuring one certain linear polarization would be enough, which is much preferable from an experimental point of view. Either way, the advantage is clear: this would require simultaneously measurement of only two Stokes parameters, which introduces only one unit of quantum noise. Moreover, if one calibrates the system with a bright signal in one of the preferred modes immediately before sending the quantum states in these modes, one could determine which Stokes parameter to measure, and then no extra units of quantum noise are added.

Since measuring only  linear polarization provides experimental advantages, a preference for PMFs preserving circular polarization is obvious for quantum information processing. These fibers are commercially available;  a typical example are spun fibers~\cite{Barlow:1981ts}. However, for those found in the market, the single-mode cutoff wavelength is greater than 1~$\mu$m, which seriously limits their applicability. Therefore, finding robust broadband polarization maintaining fiber solutions would be of great advantage for many applications. 

In this paper we want to set the stage for encoding nonorthogonal polarization states in a fiber without having to pay the quantum noise penalty.  To this end, we consider a special type of fiber, i.e., a helically twisted photonic crystal fiber (PCF), which has been successfuly fabricated in our laboratory and meets the above-mentioned requirements. Theoretical analysis has already demonstrated the favorable capabilities of these fibers~\cite{Beravat:2016aa,Edavalath:2017aa,Sopalla:2019ux}. Here, we report on the experimental demonstration of their ability to preserve circular polarization with the required precision and on the new opportunities open for improving quantum protocols.

\section{Helically twisted photonic-crystal fiber}

PMFs usually work based on the principle of optical retardance. The geometry and layout of the fiber help in maintaining certain polarization states. For instance, commercially available PMFs can maintain only linear polarization states polarized along one of the principal axes of the fiber. This is done, e. g., by introducing stress bars along one lateral axis of the fiber during fabrication to induce a difference in the refractive index. This accentuates the relevance of the alignment of the incoming polarization with respect to the PMF axes. 

Another kind of PMF is the spun optical fiber, which introduces circular retardance in the fiber through the use of stress bars spiraling along the core in propagation directions~\cite{Ulrich:1979uo,Varnham:1983wr,Barlow:1981vi}. As the name suggests, this fiber maintains circular polarization states while systematically rotating linear polarization, depending on the circular retardance. As already heralded before, the single-mode cutoff wavelength of commercially available spun fibers is greater than 1~$\mu$m, which limits their applicability. 

\begin{figure}[b]
  \centerline{\includegraphics[width=0.92\columnwidth]{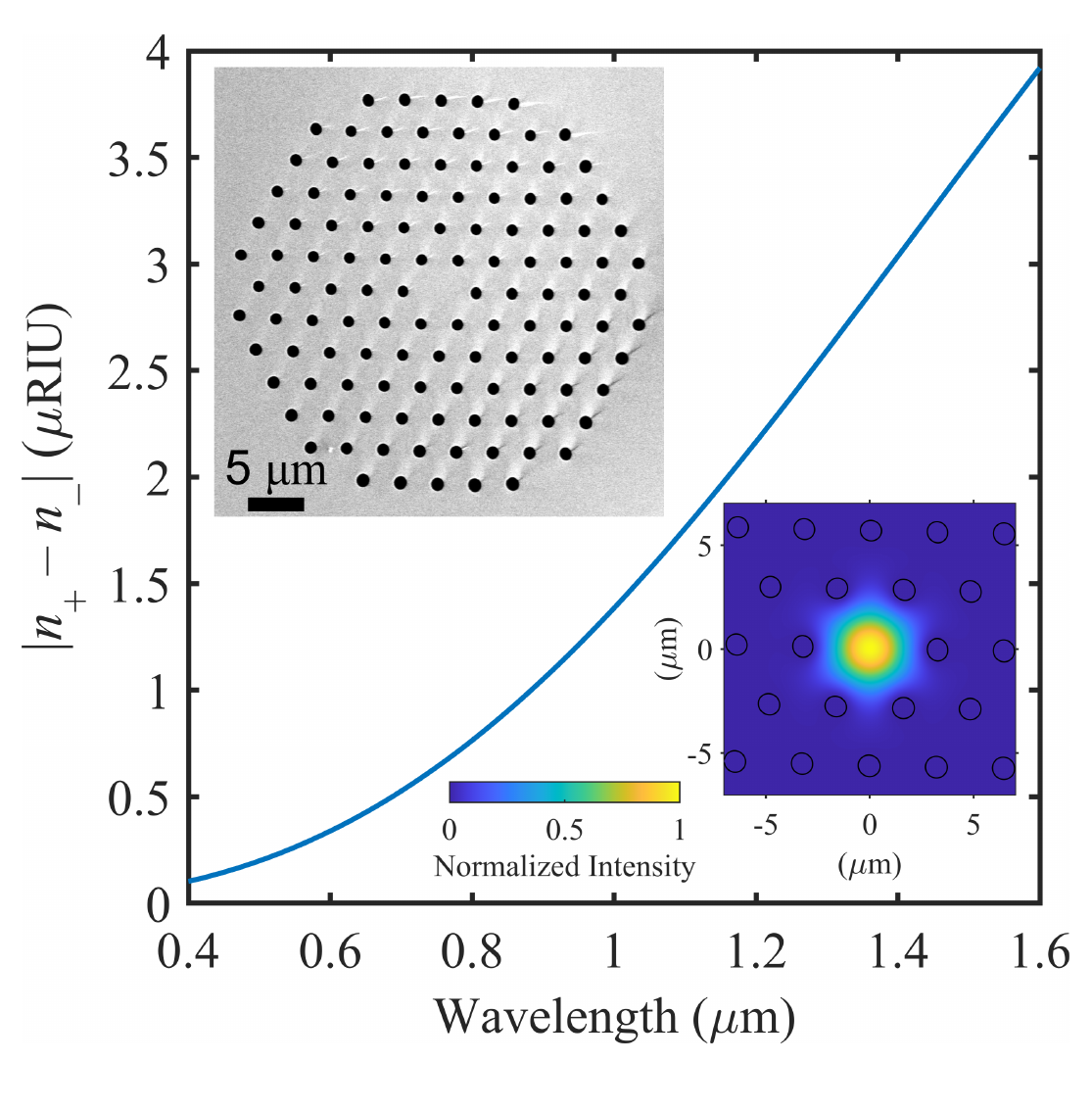}}
  \caption{Circular birefringence as a function of wavelength for the twisted PCF, calculated by finite-element modelling based on the scanning electron micrograph (SEM) dimensions. Insets: (left) SEM of the twisted PCF microstructure; (right) calculated mode field profile at 808~nm.}
  \label{fig:fiber}
\end{figure}
\begin{figure*}[t]
  \centerline{\includegraphics[width=1.75\columnwidth]{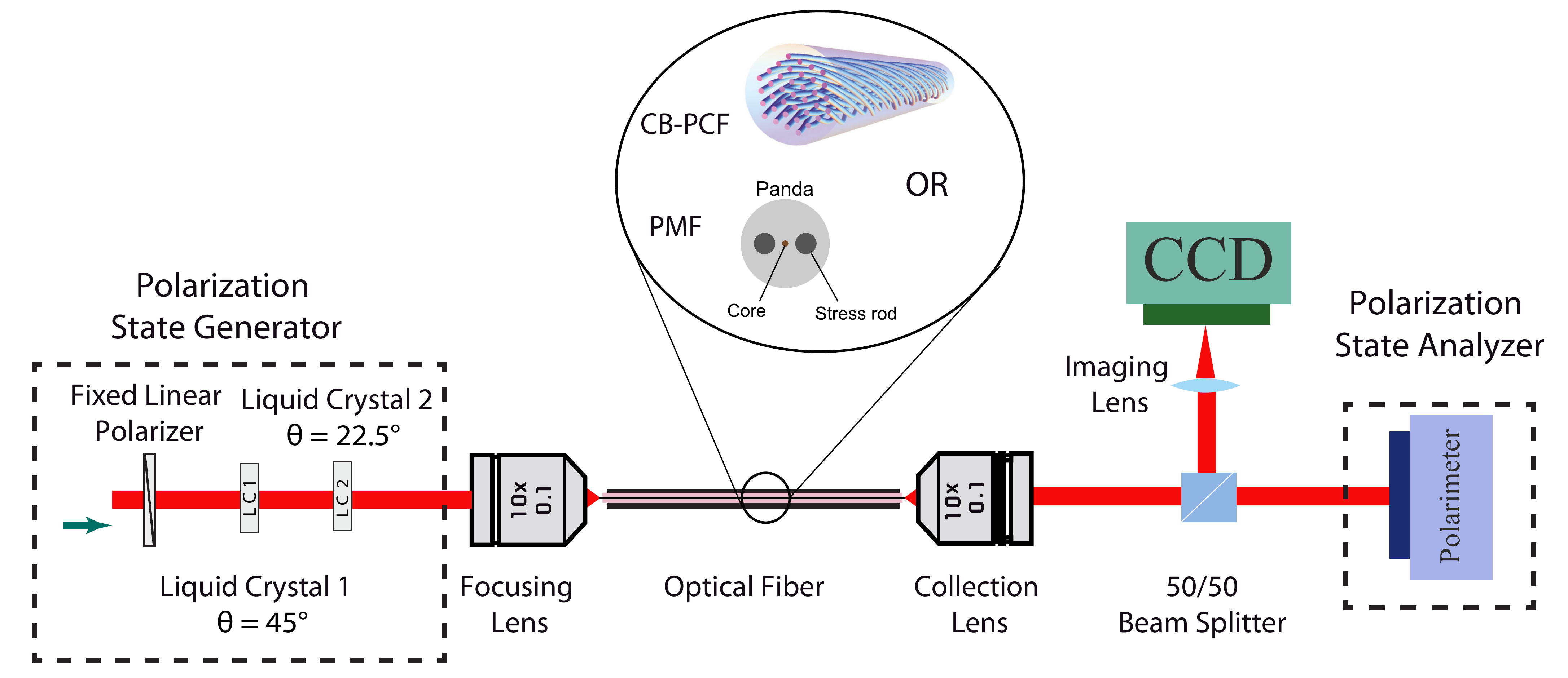}}
  \caption{Experimental setup used for polarimetric analysis of the twisted PCF and PMF. Different input polarization states are generated by the polarization state generator, using a fixed polarizer and liquid crystal variable retarders (LCVRs). Via an applied voltage, the retardance of the LCVRs can be controlled independently and without changing the in-coupling alignment. An aspheric focusing lens of numerical aperture (NA) 0.1 is used to couple light into the fiber. The transmitted light, which has traveled a distance l along the fiber, is then collected using a second lens of NA 0.1. A convex lens and camera with an 8-bit dynamic range are used to image the fiber end. This helps in making sure that light is guided through the core of fiber. The Stokes vectors of input and output polarization states are recorded using a commercial polarimeter.}
  \label{fig:setup}
\end{figure*}

The emergence of PCFs had, and is continuing to have, a tremendous impact on the field of photonics by offering enhanced modal confinement. It also offers a widely engineerable group-velocity dispersion, larger range of birefringence, and the possibility of single-mode guidance over a broad wavelength range, opening the door to alternative applications in telecommunications and beyond~\cite{Russell:2006va}. Recently, twisted PCFs (see Fig.~\ref{fig:fiber}) were introduced, providing additional advantages, such as optical activity and coreless guidance~\cite{Russell:2017tp}.

The twisted PCF in our case is fabricated from fused silica using the standard stack-and-draw technique, and the twist (period 3.6~mm) is created by spinning the preform during fiber drawing. A SEM of the fiber microstructure is shown in the top left inset of Fig.~\ref{fig:fiber}. It consists of a regular hexagonal array of hollow channels with a diameter $d = 1.6~\mu$m and interhole spacing $\Lambda = 5.2~\mu$m. Since $d/\Lambda < 0.43$, the fiber is single mode over a very broad wavelength region. The fiber outer diameter is~240 $\mu$m, and the axis of rotation coincides with the center of the fiber.

The circular birefringence $\Delta n_{\mathrm{CB}} = |n_{+} - n_{-}|$ is shown in Fig.~\ref{fig:fiber} as a function of wavelength. It has been computed in {\sc comsol} Multiphysics using the finite-element method to solve Maxwell's equations in a helicoidal coordinate system~\cite{Nicolet:2008aa} based on the SEM micrograph of the fiber microstructure. The calculated $\Delta n_{\mathrm{CB}}$ is $8 \times 10^{-7}$  refractive index unit (RIU) at 808~nm. The calculated mode field profile at 808~nm is also shown in the top right inset of Fig.~\ref{fig:fiber}. It has an effective area $A_{\mathrm{eff}}=$ 19.2~$\mu$m$^2$. For comparison, a commercial panda-type single-mode PMF is also studied in the experiment. It has a 4.5$~\mu$m core diameter and the single-mode cutoff wavelength is 700~nm. The fiber has two round stress elements, which are positioned on opposite sides of the fiber core. This induces stress along one lateral direction of the fiber, resulting in a linear birefringence $\Delta n_{\mathrm{LB}}$ in the order of $10^{-4}$.

\section{Polarimetric analysis}

To understand the optical properties of the two fibers, we perform a Mueller matrix measurement~\cite{Bass:2010vv}. An input light beam of homogeneous polarization is transmitted through the fiber under study. The polarization state of the transmitted light is a function of the fiber optical properties and the input polarization state. Input and output polarization states are defined by Stokes vectors $\mathbf{S}_{\mathrm{in}}$ and $\mathbf{S}_{\mathrm{out}}$, respectively ~\cite{Azzam:2016ua}.  Each Stokes vector consists of four components $\mathbf{S} = (S_{0}\; S_{1} \;  S_{2} \;  S_{3})^{\top}$ (the superscript $\top$ denotes the transpose),  with 
\begin{eqnarray}
S_{0} & = I_{H} + I_{V} \, ,  \qquad 
S_{1} & = I_{H} - I_{H} \, , \nonumber \\
\\
S_{2} & = I_{D} - I_{A} \, , \qquad
S_{3} & = I_{+} - I_{-} \, , \nonumber 
\end{eqnarray}
where $I$ is the intensity of light projected onto the corresponding polarization states with horizontal and vertical ($H/V$), diagonal and antidiagonal ($D/A$) or left- and right-handed circular ($+/-$) polarization.

The output and input states are related by the linear relation~\cite{Azzam:2016ua}   
\begin{equation}
\mathbf{S}_{\mathrm{out}} = M \mathbf{S}_{\mathrm{in}} \, ,
\end{equation}
where the $4\times 4$ Mueller matrix $M$, with elements $m_{ij}$, contains information about the optical properties of the fiber under investigation, such as attenuation, diattenuation and birefringence, which can be extracted by using decomposition techniques~\cite{,Lu:1996uv,Noble:2012vb,Arteaga:2010vh,Simon:2010wu}. For an unambiguous determination of the Mueller matrix, different input polarization states and their respective output states must be recorded. The projected polarization states can be recorded by various techniques such as  a combination of a polarizer and a rotating quarter-wave plate (QWP) or by using a fixed polarizer together with liquid-crystal variable retarders~\cite{Bueno:2000wt}.

\begin{figure*}[t]
  \centerline{\includegraphics[width=1.55\columnwidth]{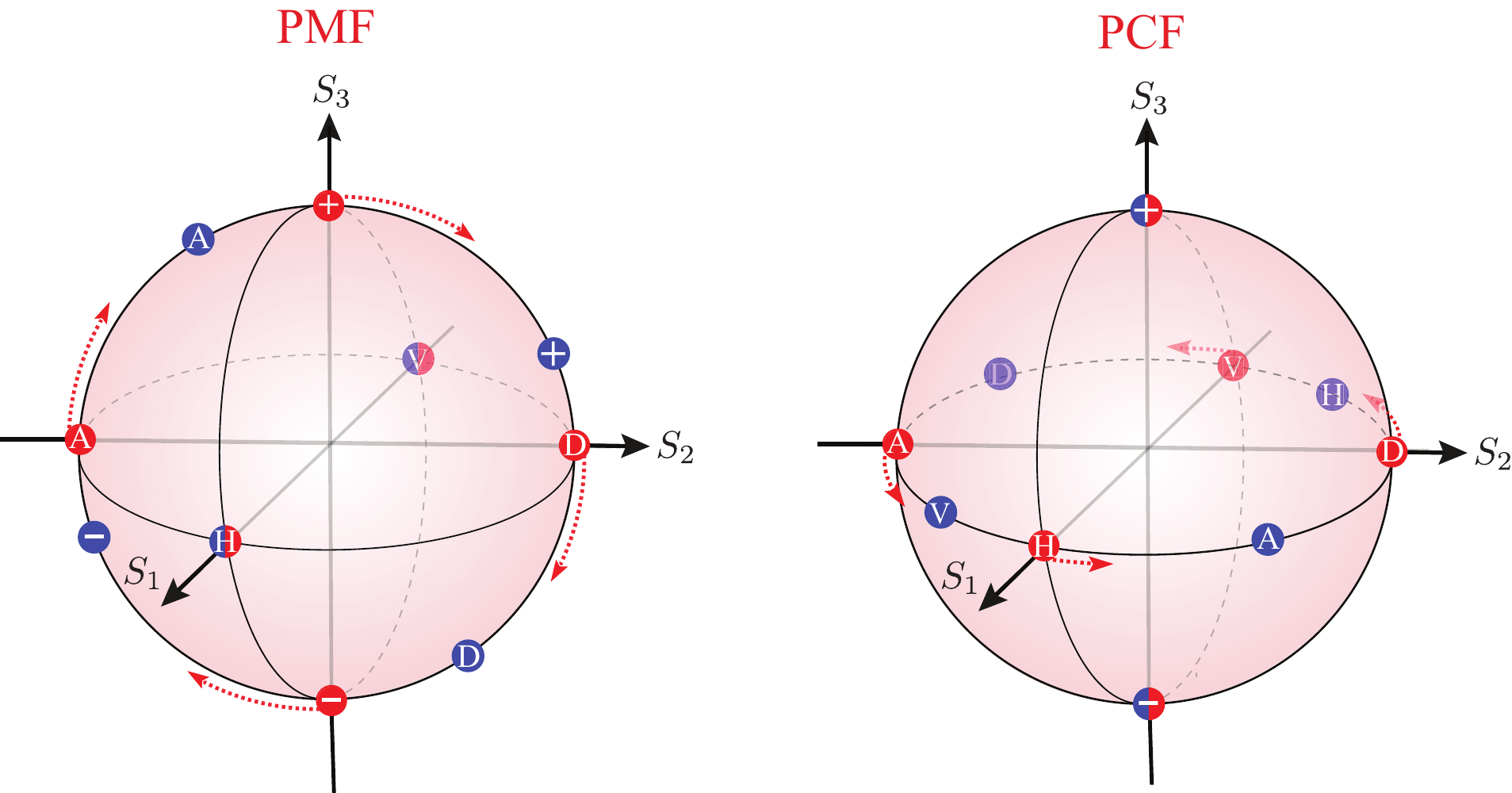}}
  \caption{Experimentally recorded polarization states (input and output) red for the PMF ($l= 5~$m, coiled; left panel) and for the twisted PCF ($l=4.6~$m, coiled; right panel). The red arrows indicate the direction of the polarization transformation.  The data shown is normalized to the length ($l$) of the fiber considering multiple retardation cycles. The two orthogonal input (output) polarization states are plotted on the Poincar\'e sphere as symbols in red (blue) circles. Right ($+$) and left ($-$) circularly polarized input states are preserved for the twisted PCF, while  H and V input polarization states are preserved for the PMF at the output.  It is also pertinent to highlight that unlike twisted PCF, the PMF is sensitive to the alignment of incoming polarization state to the fiber axis. Any deviation results in polarization beating at the output.}
  \label{fig:Multvar}
\end{figure*}

For our experimental setup, sketched in Fig.~\ref{fig:setup}, a laser emitting at a wavelength of 808~nm is used. The light is coupled into the fiber under test. The output light is collimated using a collection lens. The Stokes vector of the laser beam before entering the fiber is set to $\mathbf{S}_{\mathrm{in}}$, while the Stokes vector of the transmitted beam after fiber is given by $\mathbf{S}_{\mathrm{out}}$. A commercial polarimeter (Thorlabs PAX5710) is used to record the complete Stokes vector before entering and after transmission through the fiber. 

To change the input polarization states, as required for a complete Mueller matrix retrieval, a fixed polarizer together with two liquid crystal variable retarders (LCVRs) are utilized. The retardance induced by the LCVRs can be controlled by applying a voltage across the cells. By using LCVRs, any beam shifts or deflection introduced by moving or rotating polarization elements are avoided, hence keeping the fiber coupling unaffected. In our technique, six different input polarization states are generated by aligning the optical axis of the liquid crystals at $45^\circ$ and $22.5^\circ$, respectively, and  using a combination of different retardances, as shown in Table~\ref{tab:table1}~\cite{Butt:2019vs}.

\begin{table}[b]%
\caption{\label{tab:table1} Generated polarization states resulting from different retardances set at LCVR-1 and LCVR-2 in our setup.}
\begin{ruledtabular}
\begin{tabular}{lcc}
\textrm{State}& \textrm{Retardance LCVR-1} & \textrm{Retardance LCVR-2} \\
& $\Theta = 45^\circ$ & $\Theta = 22.5^\circ$ \\
\colrule
H & $\lambda$  & $\lambda$  \\
V & $\lambda/2$ & $\lambda$ \\
D & $\lambda$  & $\lambda/2$  \\
A & $\lambda/2$  & $\lambda/2$  \\
$+$ & $\lambda/4$  & $\lambda$  \\
$-$ & $3\lambda/4$  & $\lambda$  \\
\end{tabular}
\end{ruledtabular}
\end{table}

\begin{table*}
\caption{\label{tab:res} Optical properties of twisted PCF and PMF. Experimental errors are enclosed in parentheses.}
\begin{ruledtabular}
\begin{tabular}{cccccc}
 & & \multicolumn{2}{c}{\textrm{Twisted PCF}}&\multicolumn{2}{c}{\textrm{PMF}}\\
& & 80~cm & 4.6~m (coiled)& 2~m & 5~m (coiled) \\ \hline
 \textrm{Circular}  & Experiment & $7.79(4) \times 10^{-7}$ & $8.18(7) \times 10^{-7}$ & $1.02(6) \times 10^{-8}$ & $1.94(10) \times 10^{-8}$ \\
 \textrm{birefringence}  & Simulation & $8.00 \times 10^{-7}$ & $8.00\times 10^{-7}$ & --- & --- \\
   &  &  &  &  &  \\ 
  \textrm{Linear}  & Experiment & $8.20(2) \times 10^{-8}$ & $1.46(4) \times 10^{-7}$ & $3.00(3) \times 10^{-4}$ & $4.18(7) \times 10^{-4}$ \\
 \textrm{retardance}  & Spec. sheet & --- & ---  & $ 3.50 \times 10^{-4}$ & $3.50 \times 10^{-4}$ 
\end{tabular}
\end{ruledtabular}
\end{table*}
We perform a numerical analysis of the experimental data to extract the Mueller matrix of each investigated fiber. The noise in the Stokes vector is calculated over 400 recorded data points. In this study, we are mainly interested in the retrieval of the circular and linear retardance of the fibers  $\delta_{\mathrm{CB}}$ and $\delta_{\mathrm{LB}}$. This information is present in the elements $m_{12}$, $m_{21}$,  $m_{23}$, $m_{32}$, derived from the coherency matrices and Stokes vector relation~\cite{Lu:1996uv,Noble:2012vb,Arteaga:2010vh,Simon:2010wu}. Consequently, we perform a polar decomposition of the extracted Mueller matrix~\cite{Lu:1996uv}. This operation helps to untangle the required optical properties by mathematical operations. Subsequently, to calculate the respective circular and linear birefringence $\Delta n$ we use the following relations
\begin{equation}
\Delta n_{\mathrm{CB}} = \frac{\delta_\mathrm{CB} \lambda}{2 \pi l} \, , \qquad \qquad
\Delta n_{\mathrm{LB}} = \frac{\delta_\mathrm{LB} \lambda}{2 \pi l} \, ,
\end{equation}
where $l$ is the length of the fiber, $\lambda$ the wavelength, and $\delta_\mathrm{CB}$ and $\delta_\mathrm{LB} $ are the circular and linear retardances, respectively, calculated from the experimental Mueller matrix. The extracted birefringence values for the two fibers are shown in Table~\ref{tab:res}. The measurements are taken for different lengths of the fibers. The results shown here are retrieved after phase wrapping. The experimental errors have been estimated using standard uncertainty propagation and the measured variances in the Stokes parameters. The table confirms a dominant linear birefringence for the PMF, while the twisted PCF shows a strong circular birefringence. 

As a result of their respective birefringences, the two fibers preserve linear  ($S_E = S_1$) and circular polarization states ($S_E=S_3$), respectively. In Fig.~\ref{fig:Multvar}, we plot the input and output states for both fibers on the Poincar\'e sphere. As shown in the right panel, the twisted PCF maintains circular states very well. The minor deviations are a consequence of imperfections in the fiber, which also results in a residual linear birefringence (see Table~\ref{tab:res}). It should be noted that, unlike PMF, the coupling of light to the twisted PCF is rotationally invariant, because it has no principal axes in the transverse plane, easing the alignment procedure drastically. We can also observe in Fig.~\ref{fig:Multvar} that a linearly polarized input state coupled into the twisted PCF is rotated along the equator of the Poincar\'e sphere ($S_1-S_2$ geodesic). 

As discussed in the Introduction, the twisted PCF preserves only the two polarization eigenstates of the fiber ($\pm$), so that $S_E=S_3$ (see Fig.~\ref{fig1}). This implies that a linearly polarized input state is not scrambled into other polarizations, such as elliptical or circular, but stays linearly polarized. The effect of the twisted PCF is merely a rotation of the polarization direction. Consequently, linearly polarized light is only scrambled along the $S_1-S_2$ geodesic. Hence, the measurement task reduces to measuring one certain linear polarization. As stressed before, this might be of paramount importance for the QKD applications since any arbitrary linear input polarization can be retrieved by evaluating output linear polarization using just an analyzer.

Many commercially available fibers for preserving circular polarization states are made by spinning linearly birefringent fibers. The presence of linear birefringence has the consequence that the spun fiber does not exhibit pure circular birefringence~\cite{Laming:1989uy}. In contrast to previous works on twist-induced circular birefringence, the PCF with sixfold rotational symmetry has zero linear birefringence, provided the structure is perfectly formed~\cite{Xi:2013vk}. The twisted PCF's pure circular birefringence and single-mode guidance over a broad wavelength range makes it very convenient for several applications~\cite{Sopalla:2019ux}. Since the twisted PCF is made from a single material (fused silica), it also provides very stable thermal performance. 

Given the experimental numbers for the different types of birefringence, one should also estimate the coupling to other modes. The period between coupling between different modes depends on the difference in the effective index, which is on the order of $B_{m} \sim 10^{-6}$. This  gives a beat length of $L_B = \lambda/B_m = 0.8$~m. The power oscillating between the modes  depends on the overlap integral between the modes \cite{OKAMOTO2022}, which in this case we can estimate to be $7.66 \times 10^{-12}$ at 808~nm. This is a low value, indicating that coupling to other modes should be very low for this fiber.

In the left panel of Fig.~\ref{fig:Multvar} it also can be seen that for the PMF naturally only linear $H$ and $V$ polarization modes (or the ones parallel to the optic axes of the fiber) are maintained, confirming that $S_E=S_1$. This again implies that the input polarization mode is not scrambled all over the sphere but only along the great circle defined by $S_E$. Hence, it is going to be scrambled in one-dimensional subspace. However, in this case, the measurement requires measuring in a certain linear combination of $S_3$ and $S_2$ (elliptical polarization) and, thus, a combination of QWP and analyzer is required. From the experimental results, it can also be inferred that the alignment of the incoming modes with respect to the optical axes of the PMFs is very crucial for maintaining polarization states, constituting a major problem in practical applications such as fiber-based communication systems. Also in systems not using polarization as the parameter for encoding information, the discussed polarization stability and alignment insensitivity can be experimentally very valuable, too.

The advantages of twisted PCFs in terms of polarization measurement and  single-mode guidance over a broad wavelength range, make them ideal candidates for versatile point-to-point QKD communication systems. It would be interesting to investigate their performance in terms of quantum, thermal, and vibration-based fluctuations. Based on those results, the current twisted PCF design could then be fine tuned to match the needs required for new quantum communication systems.

\section{Concluding remarks}

In summary,  we have shown that PMFs are advantageous for transmitting quantum states in certain modes, in particular if these modes are not the preserved ones. The advantage is demonstrated for twisted PCFs because this eases the experiment. For this purpose, we have experimentally investigated the optical properties of twisted PCFs with regard to their suitability for quantum communication protocols. We have demonstrated that they have the ability to maintain circular polarization with the required fidelity and to systematically rotate linear polarization modes, leading to potential applications involving polarization-encoding schemes. This is in contrast to standard PMFs, which preserve certain linear polarization modes (if properly aligned with their two main axes) and also limit the scrambling to a great circle, but which have the disadvantage that on this great circle one finds all types of polarization:  linear, elliptical,  and circular. This is a clear disadvantage for the experimental implementation. 

Moreover, twisted PCFs are rotationally invariant to incoming polarization states, further easing their utilization in real-world setups. Together with their broadband operation and the stable thermal performance, these advantages render them very useful for robust polarization-based applications. However, it should be noted that also in systems utilizing encoding methods other than polarization, the studied polarization stability and alignment insensitivity can be very advantageous. These special properties of twisted PCFs might be highly beneficial for QKD applications.

\acknowledgments
This work is supported by the German Federal Ministry of Education and Research (BMBF) in the framework of the project QuNet and by the German Research Foundation (Deutsche Forschungsgemeinschaft, DFG) by funding the Erlangen Graduate School in Advanced Optical Technologies (SAOT) within the German Excellence Initiative. LLSS acknowledges financial support from the Spanish Ministerio de Ciencia e Innovaci\'on (Grant No PID2021-127781NB-I00). 

%

\end{document}